\documentclass{emulateapj}
\usepackage{graphicx}
\def\gsim{\;\lower4pt\hbox{${\buildrel\displaystyle >\over\sim}$}\;}
\def\lsim{\;\lower4pt\hbox{${\buildrel\displaystyle <\over\sim}$}\;}
\def\grls{\;\lower4pt\hbox{${\buildrel\displaystyle >\over <}$}\;}

\begin{document}
\title{Magnetic Sunyaev-Zel'dovich effect in galaxy clusters}
\author{Jian Hu\altaffilmark{1} and Yu-Qing Lou\altaffilmark{1,2,3}}
\altaffiltext{1}{Physics Department and the Tsinghua Center for
Astrophysics (THCA), Tsinghua University, Beijing 100084, China. }
\altaffiltext{2}{Department of Astronomy and Astrophysics,
The University of Chicago, 5640 South Ellis Avenue, Chicago,
IL 60637, USA.}
\altaffiltext{3}{National Astronomical Observatories,
Chinese Academy of Sciences, A20, Datun Road, Beijing
100012, China. }


\begin{abstract}
This Letter explores influences of intracluster magnetic fields
($\gsim 1\mu$G) submerged in the hot electron gas on classic
Sunyaev-Zel'dovich effect (SZE) and thermal bremsstrahlung in
X-ray emissions. As the Larmor frequency is much higher than all
collision frequencies, the presence of magnetic field may lead
to an anisotropic velocity distribution of hot electrons. For the
two-temperature relativistic Maxwell-Boltzmann distribution, we
compute modifications to the classical thermal SZE. Intracluster
magnetic fields tend to enhance the SZE with steeper radial
variations, which bear important consequences for cluster-based
estimates of cosmological parameters. By applying the magnetic
SZE theory to spectral observations of SZ and Chandra X-ray
emissions from the galaxy cluster Abell 2163, a $\sim 30-40\mu$G
central core magnetic field $B_0$ is predicted. For the SZ and
Chandra X-ray spectral observations of the Coma cluster, our
theoretical analysis is also consistent with an observationally
inferred $B_0\lsim 10\mu$G. As the magnetic SZE is redshift $z$
independent, this mechanism might offer a potentially important
and unique way of probing intracluster magnetic fields in the
expanding universe.
\end{abstract}


\section{Introduction}

The Sunyaev-Zel'dovich effect (SZE) in galaxy clusters offers a
unique and powerful observational tool for cosmological studies.
There has been persistent progress in detecting and imaging the
SZE in clusters. In view of this rapid development in SZE
observations, several important physical effects associated with
the classical thermal and kinetic SZEs (Sunyaev \& Zel'dovich
1969, 1980) have been further explored for their diagnostic
potentials, such as relativistic effects (Rephaeli 1995), the
shape and finite extension of a galaxy cluster with a polytropic
temperature (Puy et al. 2000), halo rotation SZE (Cooray \& Chen
2002; Chluba \& Mannheim 2002), Brillouin scattering
(Sandoval-Villalbazo \& Maartens 2001), early galactic winds
(Majumdar et al. 2001) and cooling flows (Schlickeiser 1991;
Majumdar et al. 2001; Koch et al. 2002).

Intracluster magnetic fields have been measured using a variety of
techniques and diagnostics, including synchrotron relics and halo
radio sources within clusters, inverse Compton X-ray emissions from
clusters, Faraday rotation measures of polarized radio sources either
within or behind clusters, and cluster cold fronts in X-ray images
(Clarke et al. 2001; see Carrili \& Taylor 2002 for a recent review).
These observations reveal that most cluster atmospheres are
substantially magnetized with typical field strengths of $\gsim 1\mu$G
and with high areal filling factors out to Mpc radii. In the cores of
`cooling flow' clusters (Eilek \& Owen 2002; Taylor et al. 2001) and at
cold fronts of merging clusters (Vikhlinin et al. 2001), magnetic fields
may gain intensities of $\sim 10-40\mu$G and thus become dynamically
important.

Magnetic fields in the intracluster gas allows for particle
acceleration processes which modify specifics of heating processes,
such that the electron energy distribution differs from the
Maxwell-Boltzmann distribution. Such stochastic acceleration
processes include shocks and magnetohydrodynamic (MHD) waves, etc.
The bremsstrahlung from a modified Maxwell-Boltzmann electron gas
might account for the observed X-ray spectra up to highest energies
of current X-ray observations (Ensslin et al. 1999; Blasi 2000a).
If energy injections by MHD waves are turned off, a galaxy cluster
gradually thermalizes with electrons approaching a Maxwell-Boltzmann
distribution on a rough timescale of $\sim 10^7-10^8$ yrs. As all
collision frequencies (Nicholson 1983) are much lower than the
electron Larmor frequency for the magnetized intracluster gas
(Sarazin 1988), the electron velocity distribution is likely
to be anisotropic as long as the parallel (relative to magnetic
field ${\bf B}$) pressure is not too much higher than the
perpendicular pressure (Parker 1958; Hasegawa 1975).

We presume the result of a partial electron thermalization
is a two-temperature relativistic Maxwell-Boltzmann
distribution, i.e. an anisotropic velocity distribution.
This two-temperature does not mean an electron gas having two
components with different temperatures, but refers to the same
population with different average kinetic energies along and
perpendicular to the local magnetic field. The main thrust
of this Letter is to advance magnetic SZE theory in contexts of
Chandra X-ray and radio SZE spectral observations for galaxy
clusters.

In Section 2, we calculate the X-ray emission and SZE spectra
using the two-temperature relativistic Maxwell-Boltzmann
distribution for electron velocity. Based on both Chandra X-ray
and SZ spectral observations, we offer a specific prediction
for the galaxy cluster A2163. Finally, we discuss cosmological
implications of our magnetic SZE theory in Section 3.

\section{Magnetic Sunyaev-Zel'dovich Effect }

Both X-ray emission and SZE spectra are sensitive to the hot
electron energy distribution. By the presence of $\mathbf B\gsim
1\mu$G, electrons thermalize their parallel and perpendicular
(relative to ${\bf B}$) kinetic energies separately with a
resulting two-temperature Maxwell-Boltzmann distribution on
timescales of $\sim 10^7-10^8$ yrs. We adopt a two-temperature
thermal relativistic Maxwellian electron velocity distribution
$p_{\rm e}(\beta_1,\beta_2)$, namely
\begin{equation}
p_{\rm e}(\beta_1,\beta_2)d\beta_1{\rm d}\beta_2\propto
\gamma^5\beta_2\exp\bigg(-\frac{\gamma_1}{\Theta_1}
-\frac{\gamma_2}{\Theta_2}\bigg){\rm d}\beta_1{\rm d}\beta_2\ ,
\end{equation}
where $\Theta_1\equiv k_{\rm B}T_{\parallel}/(m_{\rm e}c^2)$,
$\Theta_2\equiv k_{\rm B}T_{\perp}/(m_{\rm e}c^2)$, $k_{\rm B}
=1.38\times 10^{-16}\hbox{ erg K}^{-1}$ is the Boltzmann constant,
$c$ is the speed of light and $m_{\rm e}$ is the electron mass;
$T_{\parallel}$ and $T_{\perp}$ are parallel and perpendicular
temperatures, respectively; $\beta_1\equiv v_{\parallel}/c$,
$\beta_2\equiv v_{\perp}/c$, $\beta^2=\beta_1^2+\beta_2^2$,
$\gamma=(1-\beta^2)^{-1/2}$ and $\gamma_i=(1-\beta_i^2)^{-1/2}$
for $i=1,2$ with $v_{\parallel}$ and $v_{\perp}$ being parallel
and perpendicular velocities, respectively. We assume $\mathbf B$
to be random over the entire cluster on scales larger than a
typical coherence length of $\sim 1-10$kpc. Thus microscopically
anisotropic electrons are macroscopically isotropic, analogous
to a demagnetized ferromagnet. Integrating in all directions,
the electron speed or energy distribution becomes
\begin{equation}
p_{\rm e}(\beta){\rm d}\beta=N\gamma^5\beta {\rm d}\beta
\int_{-\beta}^{\beta}\exp\bigg(-\frac{\gamma_1}{\Theta_1}
-\frac{\gamma_2}{\Theta_2}\bigg){\rm d}\beta_1\ ,
\end{equation}
where $N$ is a normalization factor computed numerically.

The firehose stability (Parker 1958; Hasegawa 1975) for
velocity anisotropy requires
\begin{equation}
B^2/(2\pi)\gsim n_{\rm e}(\langle mv_\parallel^2\rangle
-\langle mv_\perp^2\rangle)>0\ ,
\end{equation}
where $n_{\rm e}$ is the electron number density,
$m\equiv\gamma m_{\rm e}$, angled brackets indicate averages
over $p_{\rm e}(\beta)$ and the mean temperature $T$ is
defined by $ 3k_{\rm B}T/2\equiv (2\langle mv_\perp^2\rangle
+\langle mv_\parallel^2\rangle)/2$. It then follows that
\begin{eqnarray}
 k_{\rm B}T_{\parallel}\equiv\langle mv_{\parallel}^2\rangle
=k_{\rm B}T+B^2/(3\pi n_{\rm e})\nonumber\ ,\\
 k_{\rm B}T_{\perp}\equiv\langle
mv_{\perp}^2\rangle =k_{\rm B}T-B^2/(6\pi n_{\rm e})\nonumber
\end{eqnarray}
with $B^2/(2\pi n_{\rm e}k_{\rm B})$ being the upper bound
for the temperature difference $T_\parallel-T_\perp$. For an
observed X-ray energy spectrum and an empirically inferred $B$
distribution in a cluster, one may estimate $T_\parallel$ and
$T_\perp$ by fitting the spectral data. By correlations between
X-ray surface brightness and Faraday rotation measure (Dolag et
al. 2001), a power law $B\propto [n_{\rm e}(r)]^{\alpha}$ was
inferred with an exponent $\alpha$ estimated from the slope of
$\ln B$ versus $\ln n_{\rm e}$ relation. For example, one finds
$\alpha\sim 0.9$ for the galaxy cluster A119 and $\alpha\sim 0.5$
for the galaxy cluster 3C 129 with a larger uncertainty in the
latter. For $\alpha=0.5$, $T_\parallel$ and $T_\perp$ can remain
constant in a magnetized galaxy cluster.

The X-ray emission rate per unit volume
per unit energy interval is given by
\begin{equation}
j_{\rm X}(E_{\rm X})=n_{\rm e}^2 c\int {\rm d}\beta p_{\rm
e}(\beta)\beta\sigma_{\rm B}(\beta ,E_{\rm X}),
\end{equation}
where $\sigma_{\rm B}(\beta,E_X)$ is the differential cross
section (Haug 1997) for the bremsstrahlung of an X-ray photon with
energy $E_{\rm X}$ from an electron of speed $c\beta$. Given a
central $B_0$, we fit an observed X-ray energy spectrum with the
electron distribution (2) and $T_\parallel-T_\perp$ constrained
by marginal firehose stabilities (Parker 1958; Hasagawa 1975).

To scatter a cosmic microwave background (CMB) photon (Birkinshaw
1999) of frequency $\nu_i$ off an isotropic distribution of
thermal electrons with speed $c\beta$, the probability for the
scattered photon of frequency $\nu_i(1+s)$ is
\begin{eqnarray}
P(s,\beta)=\frac{3}{16\gamma^4\beta}\int^{\mu_2}_{\mu_1}
\frac{(1+\beta\mu')}{(1-\beta\mu)^{3}}
\qquad\qquad\qquad\qquad\nonumber\\
\times[1+\mu^2\mu'^2+(1-\mu^2)(1-\mu'^2)/2]{\rm d}\mu\ ,
\quad
\end{eqnarray}
with $\mu'\equiv [(1+s)(1-\beta\mu)-1]/\beta$, where
$\mu_1=(s-\beta)/[(1+s)\beta]$ and $\mu_2=1$ for $s\geq 0$ and
$\mu_1=-1$ and $\mu_2=(s+\beta)/[(1+s)\beta]$ for $s<0$. As the
intracluster electron gas is macroscopically isotropic and has a
thin optical depth $\tau_{\rm e}\sim 10^{-2}$, integral (5) is
applicable to the magnetic SZE analysis. For photons scattered by
an electron distribution of expression (2), the resulting
distribution in the fractional frequency shift $s$ is
\begin{equation}
P_1(s)=\int^1_{s/(2+s)}{\rm d}\beta p_{\rm e}(\beta)P(s,\beta)\ .
\end{equation}
For CMB photons scattered by a hot intracluster
electron gas, the change in the CMB spectrum at
frequency $\nu$ caused by the magnetic SZE is
\begin{equation}
\Delta I(\nu)={2h\nu^3\over c^2}\tau_{\rm e}
\int^{+\infty}_{-1}\!\!\!\!\!{\rm d}s\bigg[{P_1(s)(1+s)^3
\over e^{(1+s)x}-1}-{P_1(s)\over e^x-1}\bigg],
\end{equation}
where $x\equiv h\nu/(k_{\rm B}T_{\rm CMB})$, $h=6.63\times
10^{-27}\hbox{erg s}$ is the Planck constant, $T_{\rm CMB}$ is the
present CMB temperature and $\tau_{\rm e}=\sigma_{\rm T}N_{\rm e}$
with $\sigma_{\rm T}=6.65\times 10^{-25}\hbox{ cm}^2$ being the
electron Thomson cross section (Rybicki \& Lightman 1979)
and $N_{\rm e}$ being the column density of free electrons
along the line of sight.

In the extensively used beta-model for intracluster hot
electron gas (Sarazin 1988; Fabian 1994), the empirical
radial distribution of $n_{\rm e}(r)$ is
\begin{equation}
n_{\rm e}(r)=n_{\rm e0}\big[1 +(r/r_{\rm c})^2\big]^{-3\beta_{\rm
c}/2}\ ,
\end{equation}
where $r_{\rm c}$ is the core radius and $n_{\rm e0}$ is the
central electron number density. It follows that $\tau_{\rm e}$
at radius $r$ is given by
\begin{equation}
\tau_{\rm e}(r)=\tau_{\rm e0}\big[1 +(r/r_{\rm
c})^2\big]^{(1-3\beta_{\rm c})/2}\ ,
\end{equation}
where $\tau_{\rm e0}=n_{\rm e0}\sigma_{\rm T}r_{\rm c}
\sqrt{\pi}\Gamma(3\beta_{\rm c}-1/2)/\Gamma(3\beta_{\rm c})$ and
$\Gamma$ is the gamma function.\footnote{Here, $\beta_{\rm c}$
instead of $\beta$ is used to avoid notation confusions with the
dimensionless thermal electron speed $\beta$.} We use eqns. (2),
(5), (6) and (9) in integral (7) to compute magnetic SZE
spectrum at $r$.

\begin{figure}
\includegraphics[angle=0,totalheight=4.5cm,width=7.0cm,scale=0.4]
{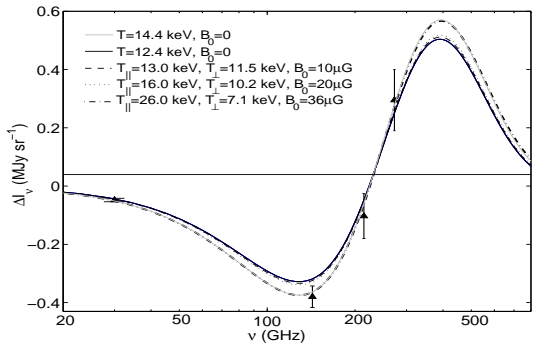} \caption{The SZE spectrum of A2163 (triangles with
error
bars are data points). The heavy solid line ($T=12.4$ keV) is the
thermal SZE with parameters determined from X-ray observations with
$B_{0}=0$. The dash, dotted and dash-dotted lines represent magnetic
SZE with different central magnetic field strengths $B_0$ and with
other cluster parameters being the same. We take $\alpha=0.5$, so
that $T_{\perp}$ and $T_{\parallel}$ remain constant in the
cluster. The grey line ($T=14.4$keV, $B_{0}=0$) coincides with
the dash-dotted line.}
\end{figure}

\begin{figure}
\includegraphics[angle=0,totalheight=5.0cm,width=8.0cm,scale=0.9]
{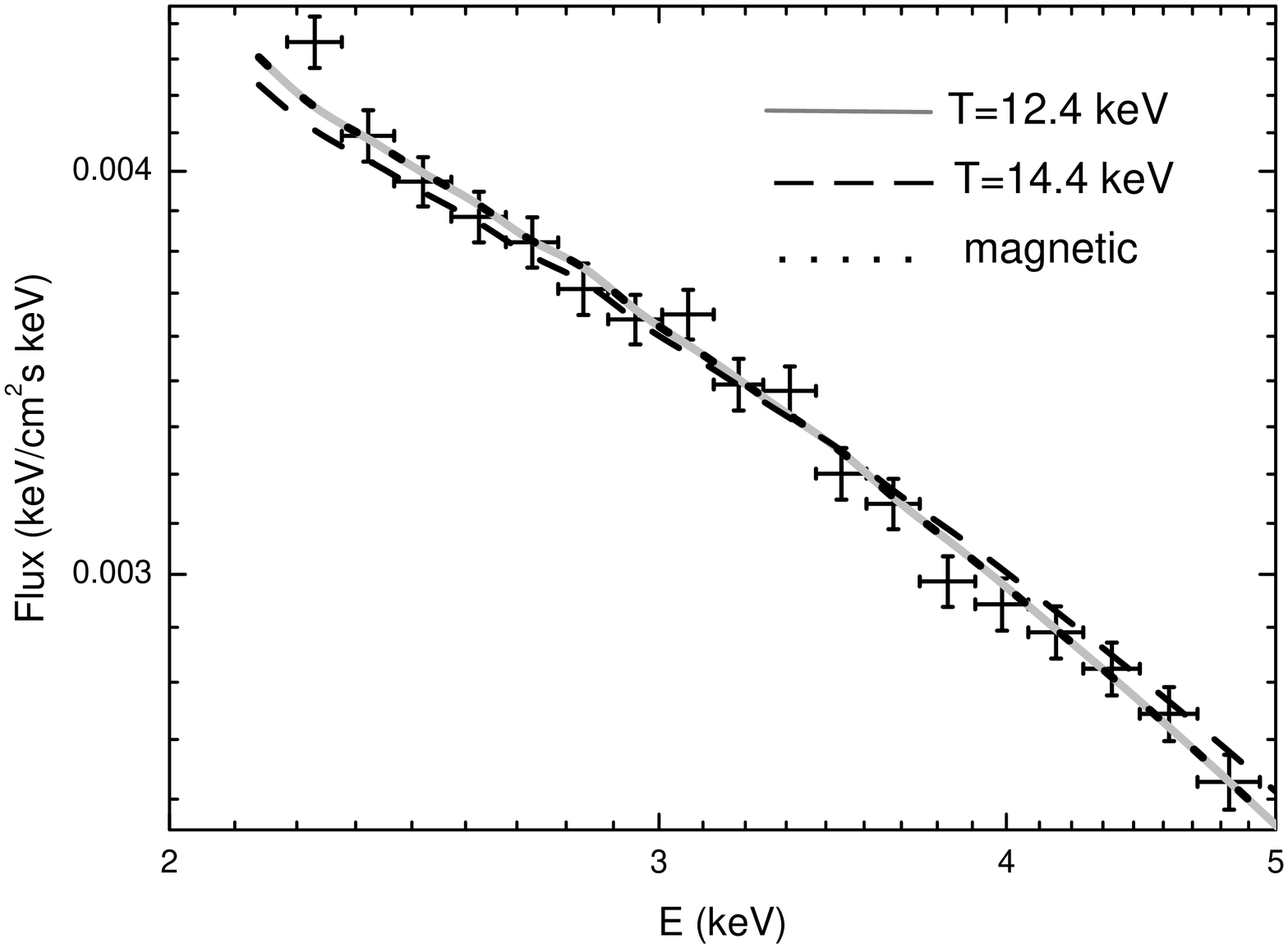} \caption{Energy spectrum of the central $r\lsim5'$
region of A2163 by Chandra. The grey and dashed lines are the
$T=12.4$ and $14.4$ keV thermal bremsstrahlung fits. The dotted
line shows our magnetic model of essentially the same fit as
$T=12.4$ keV yet with different $B_0$, $T_{\parallel}$ and
$T_{\perp}$ of Fig. 1. The lower energy part ($\lsim 2.1$ keV) is
contaminated by the soft excess (Markevitch \& Vikhlinin 2001).
The spectral resolution for energy $\gsim 5.0$ keV is not high
enough and is blended with the Fe K$_\alpha$ line (peaked at 5.3
keV for $z=0.203$). We fit the $2.1-5.0$ keV band with
uncertainity in $T$ of 0.9 keV (90\%CL). }
\end{figure}

\begin{figure}
\includegraphics[angle=0,totalheight=4.5cm,width=8.0cm,scale=0.4]
{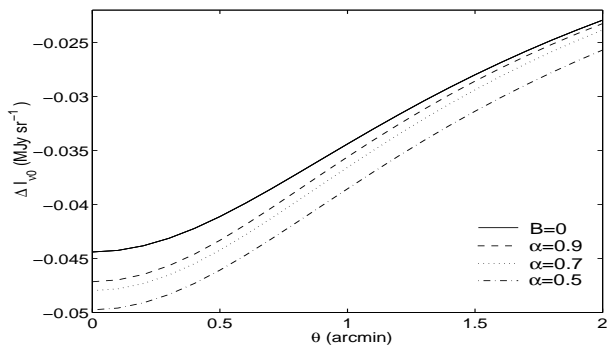} \caption{Radial features of thermal and magnetic
SZEs modeled for A2163 at $\nu_0=30$GHz. The solid line is the
thermal SZE, the dash, dotted and dash-dotted lines are models
with the same central magnetic field strength $B_0=36\mu $G but
different $\alpha$ values. }
\end{figure}

For spectral observations of A2163, we estimate the lower limit of
intracluster $B$ from data using our magnetic SZE theory. Being
one of the hottest clusters, the thermal electron gas trapped in
A2163 (redshift $z=0.203$) has a mean temperature $k_{\rm B}T_{\rm
c}\simeq 12.4\pm 0.5$keV and a central density $n_{\rm e0}\simeq
6.82\times 10^{-3} {\rm cm}^{-3}$ with a core radius $r_{\rm c}=
0.269\pm 0.025$Mpc (Hubble constant $h_{100}=0.71$) and a
$\beta_{\rm c}=0.616\pm 0.031$ (Elbaz et al. 1995; Markevitch et
al. 1996; Markevitch \& Vikhlinin 2001). By these estimates, the
optical depth towards the cluster center is $\tau_{\rm e0}=0.0133$
and the central Compton parameter is $y_{\rm th0}=3.21\times 10^{-4}$.

A conspicuous SZE spectrum has been observed in A2163 at four
frequencies by BIMA at 30GHz (LaRoque et al. 2002), by DIABOLO
at 140GHz (Desert et al. 1998) and by SuZIE at 140, 218 and
270GHz (Holzapfel et al. 1997) with dust-corrections (LaRoque
et al. 2002). These data were previously fit with a Compton
parameter $y_{\rm th}=(3.56^{+0.41+0.27}_{-0.41-0.19})\times
10^{-4}$ for a thermal SZE together with a kinetic SZE for a
positive peculiar velocity
$V_{\rm p}=415_{-850\ -440}^{+1030+460}\ {\rm km\ s^{-1}}$
(68\%CL) (Carlstrom et al. 2002). A positive-velocity kinetic
SZE leads to an overall downward shift of the thermal SZE
spectrum especially around the zero point ($\sim218$ GHz)
of the SZE.

We fit the observed X-ray spectrum of Fig. 2 for a mean
temperature $T\cong 12.4$keV by eqns. (2) and (4) with different
$B$ strengths and infer relations among $B,\  T_\parallel
\hbox{ and }T_\perp$. We here take $\alpha=0.5$ with $T_{\perp}$
and $T_{\parallel}$ being constants (see parameters of Fig. 1).
Inserting these relations into eqns. (2), (6) and (7), we obtain
the spectra of magnetic SZE.

Shown in Fig. 1 are SZE spectra computed for A2163 with different
parameters in comparison with the observed SZE spectrum. The model
without magnetic field (solid line for $T=12.4$keV) underestimates
signals of SZE (especially at $\nu=140$GHz) with $\chi^2=3.82$
in a chi-square fit. Intracluster magnetic field tends to enhance
signals of SZE and the best fit is $B_0=36\mu$G with $\chi^2=0.78$.
As the marginal firehose stability (Parker 1958) is used here, the
fitting estimate represents the lower limit of $B_0=36\mu$G for
A2163. Magnetic field also increases the null frequency of the
pure thermal SZE, similar to the kinetic SZE with a positive
velocity. This degeneracy can be removed by SZE signals at other
frequencies (e.g. 100 and 400GHz, etc.). The SZE spectrum data
can be fit with $T=14.4$keV (grey line in Fig. 1) and $B_0=0$,
but the thermal spectrum data does not fit well with $T=14.4$keV
(dashed line in Fig. 2). Shown in Fig. 3 are computed radial SZE
features of thermal ($B=0$) and magnetic ($\alpha\neq 0$) SZEs
modeled for cluster A2163 at 30GHz. With a smaller exponent $\alpha$,
SZE signals steepen from the central to peripheral parts of a cluster.
This can be utilized to determine macroscopic mean $\mathbf B$
structures by obtaining spatially resolved magnetic SZE intensity
maps (Carlstrom et al. 2002). Note that 30GHz of Fig. 3 is just an
illustrating example.

\section{Discussions }

Contrary to recent results (Koch et al. 2003; Zhang 2003), we
find that the anisotropic velocity distribution of electrons
caused by magnetic field $B$ enhances the SZE. Our model results
of Figs. $1-3$ can be critically tested against more precise
spectral SZE measurements of A2163 in the frequency bands of
$\sim 50-130$GHz and $\sim 300-600$GHz by MAX,
MSAM
and SuZIE
types of experiments in the frequency passbands of $90-670$GHz and
by AMiBA in the band $84-104$GHz, Nobeyama at 21 and 43GHz, JCMT
at 350 and 650GHz, SZA
in the bands of $26-36$GHz and $85-115$GHz, BIMA and OVRO in the
band of $26-36$GHz,
MINT
at 150GHz and ACT
at 150, 220, and 270 GHz. Multi-frequency projects such as the
upgraded MITO (Lamagna et al. 2002; De Petris et al. 2002) and
the OLIMPO (Masi et al. 2003) experiments are very promising to
provide some results.

A2163 may involve merger shocks that could amplify $B$ (e.g.
Markevitch \& Vikhlinin 2001). The spectral index maps of A2163
show a spectral steepening from the central to peripheral radio
halo regions, implying a radial decrease of $B$ in reacceleration
models (e.g. Feretti et al. 2003). It was attempted to fit the
SZE spectrum of A2163 with a combination of thermal and non-thermal
electrons (e.g. Colafrancesco et al. 2003), but no evidence was
found for hard X-ray excess due to the non-thermal component in
the BeppoSAX data (e.g. Feretti et al. 2001).

Based on X-ray and SZE measurements, 41 galaxy clusters were used
to independently estimate Hubble constant $h_{100}=0.61\pm 0.03\pm
0.18$, where the uncertainties are statistical and systematic at
68\% confidence level for $\Omega_M=0.3$ and
$\Omega_{\Lambda}=0.7$ cosmology (Carlstrom et al. 2002; Reese 2003).
Our analysis of A2163 shows that intracluster magnetic field induces
microscopic anisotropies in electron velocity distribution to enhance
the SZE. It appears that inferences from cluster models without magnetic
field would systematically
underestimate $h_0$ as in the case of A2163 for which Holzapfel et
al. (1997) inferred a lower $h_{100}=0.60\pm 0.04$ against the
current WMAP result of $h_{100}=0.71\pm 0.04$. As the cluster
asphericity and orientation in the sky are random and the average
cluster peculiar velocity is zero, these factors should contribute
to the systematic uncertainty with the Hubble constant being
statistically unaltered. The underestimation of Hubble constant
may be explained by the generic presence of core magnetic field
$B_0\sim 10-40\mu$G in this sample of galaxy clusters.

Another important cosmological effect of the ubiquitous enhancement
of magnetic SZE due to the prevalence of $\gsim 1\mu$G magnetic
fields in galaxy clusters would be observable in the CMB angular
spectrum especially at high $l\gsim 3000-4000$. This contribution
to CMB fluctuations may be estimated and tested by CMB experiments
such as ACT, Planck, SZA, etc. Details of these two cosmological
effects will be pursued in forthcoming papers.

For X-ray (Arnaud et al. 2001) and SZE (De Petris et al. 2002)
spectral observations of Coma cluster (Abell 1656), our magnetic
SZE analysis is consistent with the currently inferred $B_0\lsim
10\mu$G (Carilli \& Taylor 2002). Likewise, magnetic SZE can be
utilized in other galaxy clusters with high-resolution and
high-sensitivity X-ray and SZE spectral observations to estimate
the lower limit of $B_0$ as well as SZE spatial features. While it
is necessary to estimate all possible corrections to the classic
SZE in order to isolate the magnetic contribution, this may be a
unique procedure to probe intracluster magnetic fields at high
redshifts $z$, at least statistically.
Finally, anisotropic distributions of nonthermal electrons
should lead to distinct magnetic SZE in radio lobes of
extragalactic jets.
\\

We thank the referee Dr. D. Puy for useful comments. This
research has been supported in part by the ASCI Center for
Astrophysical Thermonuclear Flashes at the U. of Chicago under
DOE contract B341495, by the Special Funds for Major State Basic
Science Research Projects of China, by the THCA, by the
Collaborative Research Fund from the NSF of China for Outstanding
Young Overseas Chinese Scholars (NSFC 10028306) at the National
Astronomical Observatory, CAS, by NSFC grant 10373009 at the
Tsinghua U., and by the Yangtze Endowment from the Ministry of
Education through the Tsinghua U. Affiliated institutions of YQL
share this contribution.



\begin{thebibliography}{}
\bibitem{}Arnaud, M., et al. 2001, A\&A, 365, L67
\bibitem{}Birkinshaw, M. 1999, Phys. Rep., 310, 97
\bibitem{}Blasi, P. 2000a, ApJ, 532, L9
\bibitem{}Blasi, P. Olinto, A. V., Stebbins, A. 2000b, ApJ, 535, L71
\bibitem{}Carlstrom, J., Holder, G., Reese, E. 2002, \araa, 40, 643
\bibitem{}Carilli, C. L., Taylor, B. 2002, \araa, 40, 319
\bibitem{}Chluba, J., Mannheim, K. 2002, A\&A, 396, 419
\bibitem{}Clarke, T. E., Kronberg, P. P., B\"ohringer, H.
2001, ApJ, 547, L111
\bibitem{}Colafrancesco, S., Marchegiani, P., Palladino, E.
2003, A\&A, 397, 27
\bibitem{}Cooray, A., Chen, X. 2002, ApJ, 573, 43
\bibitem{}De Petris, M. et al. 2002, ApJ, 574, L119
\bibitem{}Desert, F. et al. 1998, New Astron., 3, 655
\bibitem{}Dolag, K. et al.
2001, A\&A, 378, 777
\bibitem{}Eilek, J.A., Owen, F.N. 2002, ApJ, 567, 202
\bibitem{}Elbaz, D., Arnaud, M., B\"oringer, H. 1995, A\&A, 293, 337
\bibitem{}Fabian, A. C. 1994, ARA\&A, 32, 277
\bibitem{}Feretti, L. et al.
2001, A\&A, 373, 106
\bibitem{}Feretti, L. et al. 2003, The Cosmic Cauldron,
25th IAU meeting
\bibitem{}Haug E. 1997, A\&A, 326, 417
\bibitem{}Hasegawa, A. 1975, Plasma Instability
and Nonlinear Effects, Springer-Verlag, Berlin
\bibitem{}Holzapfel, W. L. et al. 1997, ApJ, 480, 449
\bibitem{}Kaiser, N. 1986, \mnras, 222, 323
\bibitem{}Koch, P. M., Jetzer, Ph., Puy, D. 2002, New Astron., 7, 587
\bibitem{}Koch, P. M., Jetzer, Ph., Puy, D. 2003, New Astron., 8, 1
\bibitem{}Lamagna, L. et al. 2002, AIP Conf. Proc. 616, 92
\bibitem{}LaRoque, S. J. et al. 2002, ApJ, submitted (astro-ph/0204134)
\bibitem{}Majumdar, S., Nath B. Chiba M., 2001, MNRAS, 324, 537
\bibitem{}Markevitch, M. et al. 1996, ApJ, 456, 437
\bibitem{}Markevitch, M., Vikhlinin A. 2001, ApJ, 563, 95
\bibitem{}Masi, S. et al. 2003, Mem.S.A.It 74, 96
\bibitem{}Nicholson, D. R. 1983, Introduction
to Plasma Theory, Wiley, New York
\bibitem{}Parker, E. N. 1958, Phys. Rev., 109, 1874
\bibitem{}Puy, D., Grenacher, L., Jetzer, Ph., Signore, M. 2000, A\&A 363, 415
\bibitem{}Reese, E. D. 2003, Measuring and Modeling the Universe,
ed. W. L. Freedman (Cambridge: Cambridge Univ. Press)
\bibitem{}Rephaeli, Y. 1995, ARA\&A, 33, 541
\bibitem{}Rosati, P., Borgani, S., Norman, C. 2002, \araa, 40, 539
\bibitem{}Rybicki, G. B., Lightman, A. P. 1979,
Radiative Processes in Astrophysics, Wiley, New York
\bibitem{}Sandoval-Villalbazo, A., Maartens, R. 2001, astro-ph/0105323
\bibitem{}Sarazin, C. 1988, X-Ray Emission from
Clusters of Galaxies, Camb. Univ. Press, Cambridge
\bibitem{}Schlickeiser, R. 1991, A\&A, 248, L23
\bibitem{}Sunyaev, R. A., Zel'dovich, Y. B. 1969, Ap\&SS, 4, 301
\bibitem{}Sunyaev, R. A., Zel'dovich, Y. B. 1980, ARA\&A, 18, 537
\bibitem{}Taylor, G. B., Fabian, A. C., Allen, S. W. 2002, MNRAS, 334, 769
\bibitem{}Vikhlinin, A., Markevitch, M., Murray, S. S. 2001, ApJ, 549, L47
\bibitem{}Zhang, P. J. 2003, astro-ph/0308354
\end{thebibliography}
\end{document}